\documentclass[10pt,conference]{IEEEtran} 
\IEEEoverridecommandlockouts

\usepackage{cite}
\usepackage{amsmath,amssymb,amsfonts}
\usepackage{algorithmic}
\usepackage{graphicx}
\usepackage{textcomp}
\usepackage{xcolor}
\def\BibTeX{{\rm B\kern-.05em{\sc i\kern-.025em b}\kern-.08em
    T\kern-.1667em\lower.7ex\hbox{E}\kern-.125emX}}

\usepackage{booktabs}
\usepackage{multirow}
\usepackage{paralist}
\usepackage{graphicx}
\usepackage{amsmath}

\usepackage{amssymb}
\usepackage[table]{xcolor}
\usepackage{xspace}
\usepackage{tikz}
\usepackage{pgfplots}
\pgfplotsset{compat=1.17}
\usetikzlibrary{arrows.meta,positioning,calc,fit,backgrounds,shapes.geometric}
\usepackage[hidelinks]{hyperref}
\usepackage{orcidlink}
\usepackage[capitalise,noabbrev]{cleveref}

\usepackage{url}
\newcommand{\sys}{\textsc{BeSpec}\xspace}

\newcommand{\pat}{Pass@1}
\newcommand{\apr}{APR}
\definecolor{specblue}{RGB}{31,119,180}
\definecolor{implorange}{RGB}{255,127,14}
\definecolor{gapred}{RGB}{214,39,40}
\definecolor{repgreen}{RGB}{44,160,44}
\definecolor{boxgray}{RGB}{240,240,240}
\newcommand{\rqconclusion}[2]{%
\begin{center}
\fcolorbox{black}{boxgray}{%
\begin{minipage}{0.92\linewidth}
\small\textbf{Answer to #1.} #2
\end{minipage}}
\end{center}}
\newtheorem{problem}{Problem} 
\begin{document}

\title{BeSpec: Behavior-Level Specification Alignment for Code Generation}

\author{%
\IEEEauthorblockN{%
Qinghua Xu\IEEEauthorrefmark{1}\,\orcidlink{0000-0001-8104-1645},
Guancheng Wang\IEEEauthorrefmark{1}\,\orcidlink{0000-0002-4338-8813},
Boxi Yu\IEEEauthorrefmark{1}\,\orcidlink{0000-0001-5213-7189},
Lionel Briand\IEEEauthorrefmark{1}\IEEEauthorrefmark{2}\,\orcidlink{0000-0002-1393-1010}}
\IEEEauthorblockA{\IEEEauthorrefmark{1}\textit{Lero the Research Ireland Centre for Software, University of Limerick}, Limerick, Ireland}
\IEEEauthorblockA{\IEEEauthorrefmark{2}\textit{University of Ottawa}, Ottawa, Canada}
\IEEEauthorblockA{\{qinghua.xu, guancheng.wang, boxi.yu\}@ul.ie, lionel.briand@lero.ie}
}
\maketitle

\begin{abstract}
Large language models (LLMs) have advanced code generation from natural-language descriptions of desired behavior (\textit{intent}). Most existing methods refine code using execution feedback while leaving the underlying specification unchanged, implicitly assuming that the LLM correctly understands the intent. Yet ambiguous or underspecified descriptions can lead even capable models to correctly implement the wrong \emph{intent}, making specification mismatch a central bottleneck.

This paper presents \sys, a behavioral model-based approach to specification alignment. Treating \emph{intent} as partial evidence of intended program behavior, \sys extracts a structured specification and predicts executable \emph{behaviors} that capture properties a correct program should satisfy. These avoid requiring exact test oracles, which can be as hard to predict as the solution itself. \sys executes candidate programs on probe inputs and compares their \textit{observed behaviors} with the \textit{predicted behaviors}. It validates mismatches and diagnoses their causes, refining defective specifications.

We evaluate \sys with three LLMs on four code-generation benchmarks: CodeContests, xCodeEval, APPS, and LiveCodeBench. Against nine baselines, \sys achieves the highest Pass@1 and average pass rate in all settings, improving mean Pass@1 over the strongest baseline by $9.3\%$--$23.6\%$ relative across LLMs. After alignment, most remaining failures are adherence failures. Ablation studies on \texttt{qwen3-coder-30b} show the benefit of behavior modeling.
\end{abstract}

\begin{IEEEkeywords}
Specification Alignment, Large Language Model, Code Generation 
\end{IEEEkeywords}

\section{Introduction}
\label{sec:intro}

Code generation aims to automatically produce source code from a programming specification, such as a natural-language problem statement (\textit{intent}). Large language models (LLMs) have made substantial progress in supporting this task. Pretrained on large and diverse corpora, they can translate \textit{intents} into executable programs and have improved the state of the art across a wide range of software engineering tasks~\cite{chen2021codex,austin2021program,li2022competition,jiang2024survey,fan2023large}. Despite their success, LLM-generated programs still often fail when requirements are complex, implicit, or only partially specified~\cite{specfix,specine,tian2023mufix,mu2024clarifygpt,reacoder,sepcalign}.

Many existing techniques improve LLM code generation by decomposing the process into stages. At a high level, the model first interprets the \textit{intent}, then plans a solution, implements the plan as code, and finally repairs the code using test feedback. Recent prompt-based and agent-based methods primarily strengthen the latter stages: they elicit better plans, coordinate specialized agents, sample diverse implementations, or use execution feedback to refine incorrect programs~\cite{chen2023selfdebug,olausson2023selfrepair,tian2023mufix,huang2023agentcoder,dong2024selfcollab,zhang2024paircoder,li2023scot}. This \textit{generate-then-refine} paradigm on the code improves pass rates and has become a dominant design pattern for LLM-based code generation. Related techniques also support feedback-guided and multi-agent test generation~\cite{mutgen,candor}.

However, improving planning, implementation, and repair is not sufficient when the first stage (i.e., intent interpretation) is wrong. Before generating any code, the LLM must infer what the task actually requires (the \textit{true intent}) from what its description literally states (the \textit{written intent}). Most refinement pipelines largely ignore this gap and treat the written \textit{intent} as correct, while, in practice, written programming \textit{intents} are often ambiguous or underspecified. For example, an \textit{intent} asking for a ``copy'' of a nested object may intend a deep copy that preserves dependencies among nested fields, while an LLM may implement a shallow copy that only duplicates the top-level object. Such ambiguities invite LLMs to fill in missing intent with their own inferred assumptions. The generated code may therefore be logically coherent and well implemented, yet still deviate from the task's true intent because it solves a slightly different problem. In this case, improving planning, implementation, or repair only helps the model implement its mistaken interpretation more effectively.
Instead, the model should reason about and clarify the specification before generating code. Prior work refers to this task as \textbf{specification alignment}~\cite{specine,specfix}, which aims to align the specification interpretation with the \textit{true intent} so that the subsequent code generation is guided by the \textit{true intent}.

A growing line of work addresses specification alignment by refining requirements before or during code generation~\cite{tian2023mufix,mu2024clarifygpt,specine,specfix}.
Among these methods, Specine~\cite{specine} and SpecFix~\cite{specfix} are the most directly related state-of-the-art baselines for our work. Specine recovers the specification implicitly inferred from the generated code and aligns it with the original input. SpecFix samples multiple candidate programs for the same ambiguous requirement, analyzes how their outputs diverge on tests, and rewrites the requirement to reduce the ambiguity that led to those divergent implementations. These methods are an important step beyond code-only repair because they make the specification itself a target of refinement.

However, existing specification alignment methods still operate mainly at the \textit{test level}. Candidate programs are executed on public or generated tests, and alignment is inferred from their final outputs, either by comparing them with expected outputs or by measuring disagreement among programs. Such test-level evidence can reveal that candidate programs disagree, or that a program fails a test, but it does not directly expose which part of the specification caused the disagreement. For example, a failed test does not tell us whether the problematic point concerns an output format rule, a feasibility condition, an indexing convention, a boundary case, or the treatment of duplicate values. As a result, test-level evidence is often too coarse to support targeted specification repair. A further limitation is that test-level alignment depends on the availability and quality of complete tests. A test must contain not only a valid and informative input, but also a correct expected output (oracle). However, in non-trivial programming scenarios, generating the correct oracle can itself be close to solving the task~\cite{Hossain2024TOGLL:LLMs,candor}. Generated tests may also be invalid, unrepresentative, or focused on cases that do not isolate the underlying ambiguity. Consequently, an incorrect oracle may make a valid program appear wrong, while an uninformative test may fail to distinguish two different interpretations of the same specification. Thus, \textit{test-level} evidence provides a noisy and indirect signal for specification alignment.

We therefore move from the \textit{test level} to the \textit{behavior level}. Rather than relying completely on tests, \sys decomposes the intended solution into smaller, checkable behavioral expectations. These behaviors capture checkable properties, such as whether the output must be exactly \texttt{YES} or \texttt{NO}, whether a solution should exist under a certain input condition, whether indices are recomputed after deletion, or how duplicate values should be handled. Unlike \textit{test-level} evidence, such behaviors do not require solving the whole task for every generated input. Therefore, they are easier to predict from the written \textit{intent}, easier to observe from candidate programs, and easier to trace back to specific vague, missing, or misread parts of the specification.  In this way, behavior-level reasoning provides a finer signal for improving the specification.

To that end, we propose \sys, a behavior-level approach to specification alignment. \sys first asks how the correct program should behave. It predicts checkable behavioral properties from the \textit{intent}, such as output constraints, feasibility conditions, and format rules.
\sys then compares these predicted behaviors with the observed behaviors of candidate programs. Note that such observed behaviors result from executing selected test inputs that have no corresponding expected outputs. After validation and diagnosis, gaps attributed to specification defects identify which specific behavioral expectation is ambiguous, underspecified, or misread. These prediction--observation gaps make specification refinement more targeted than relying solely on coarse, test-level feedback from comparing actual and expected outputs.

We evaluate \sys across three LLMs and six evaluation settings spanning four benchmarks, including the more recent LiveCodeBench. \sys achieves the highest \pat{} and  average passing rate (\apr{}) across all settings, outperforming nine baselines. Averaged over the six settings, \sys improves \pat{} over the strongest baseline by $9.3\%$--$23.6\%$ relative and improves \apr{} by $8.9\%$--$25.6\%$ across the three LLMs. The gains also hold on LiveCodeBench, where \sys improves \pat{} by $15.8\%$--$17.7\%$ relative. A failure analysis further shows that, after alignment, most remaining failures are adherence failures, primarily algorithmic/logic errors, runtime crashes, and timeouts.

This paper makes the following contributions:
\begin{compactitem}
  \item \textbf{Behavior-level specification alignment.} We advance specification alignment from coarse test-level evidence to behavior-level reasoning, where checkable behavioral expectations are predicted, observed, and traced back to vague, missing, or misread parts of the specification.
  \item \textbf{The \sys method.} We instantiate this formulation with a concrete pipeline that predicts behavioral expectations, compares them with candidate executions, automatically repairs the specification, and selects the best-aligned program.
  \item \textbf{A comprehensive evaluation.} Across three LLMs and six evaluation settings spanning four benchmarks, including the more recent LiveCodeBench, \sys outperforms nine baselines on both \pat{} and APR.
  \item \textbf{A failure analysis.} We conduct a failure analysis using an LLM-as-judge and find that, after alignment, most remaining failures are adherence failures, primarily algorithmic/logic errors, runtime crashes, and timeouts.
\end{compactitem}

\section{A Motivating Example}
\label{sec:motivation}

We adapt a programming problem from the CodeContests dataset~\cite{li2022competition} as a running example. The original \textit{written intent} is brief:
\begin{problem}[Erase Game]
You are given an integer sequence $a_1,\dots,a_n$. Repeatedly choose an index $i$
such that $a_i$ is not divisible by $i+1$, and erase $a_i$. Determine whether it
is possible to erase the entire sequence.
\end{problem}

The answer is printed as \texttt{YES} or \texttt{NO}. In this benchmark instance, the \textit{intent} is accompanied by sample cases, such as $[1,2,3]$ (\texttt{YES}) and $[7,7]$ (\texttt{YES}). These sample cases help confirm the answer format and rule out interpretations that contradict the demonstrated behavior. We include them in the motivating example because they are available in this benchmark instance and are required by several baselines. In practical settings, such sample cases may be provided by programmers together with the task or may already exist as an initial test suite, for example, in a test-driven development setting. However, they are not compulsory for \sys{}. They are useful when available, but \sys{} focuses on behavior-level reasoning rather than test-level reasoning.

One ambiguity in this \textit{intent} is whether indices are recomputed after each deletion or whether an element keeps its original index. The intended interpretation is dynamic: after an erase, the remaining sequence closes up and is indexed again from the front. This can be inferred from the examples or language conventions, but it is not stated directly. The sequence $[1,3]$
shows why the distinction matters:
\begin{compactitem}
  \item \textbf{Dynamic indices.} Erase $1$ first, since $1$ is not divisible by
  $1+1=2$. The remaining sequence is $[3]$, whose only element is now at index $1$;
  since $3$ is not divisible by $2$, it can also be erased. The answer is
  \texttt{YES}.
  \item \textbf{Original indices.} Erase $1$ first, but keep the remaining element
  $3$ at its original index $2$. It is then checked against $2+1=3$, and since $3$ is divisible by $3$, it cannot be erased. The answer is \texttt{NO}.
\end{compactitem}
Thus, the same statement can induce two plausible executions unless the reindexing rule is made explicit. A refinement loop on the code may
keep improving a candidate program while never noticing that it follows the wrong indexing convention.

\sys{} exposes this gap by comparing predicted behavior with observed candidate
behavior. From the \textit{written intent}, it predicts that after each deletion,
the remaining sequence should be reindexed. It then runs candidate programs on input scenarios and checks whether their observed behavior matches this
prediction. If all candidates behave consistently with the predicted behavior, no repair is needed. Otherwise, after validating the prediction and attributing the mismatch to a specification defect, \sys{} repairs the specification by explicitly stating the reindexing rule.

This example illustrates why behavior-level signals are more informative than test-level feedback alone. A candidate that keeps original indices may still pass available samples or fail only on a coarse input--output test, without revealing that the specific missing behavior is reindexing after deletion. By predicting the intended reindexing behavior and checking candidate executions against it, \sys{} turns this hidden interpretation gap into targeted evidence for refinement. This behavior-level design is also important in practice because public examples or labeled tests may be sparse or unavailable. In such cases, \sys{} can still reason over checkable behavioral expectations rather than depending solely on complete test cases with expected outputs. We revisit this example throughout our description of  \sys (\Cref{sec:method}).

\section{The \sys Method}
\label{sec:method}

\subsection{Overview}
\label{sec:method:overview}

\begin{figure*}[t]
  \centering
  \includegraphics[width=\textwidth]{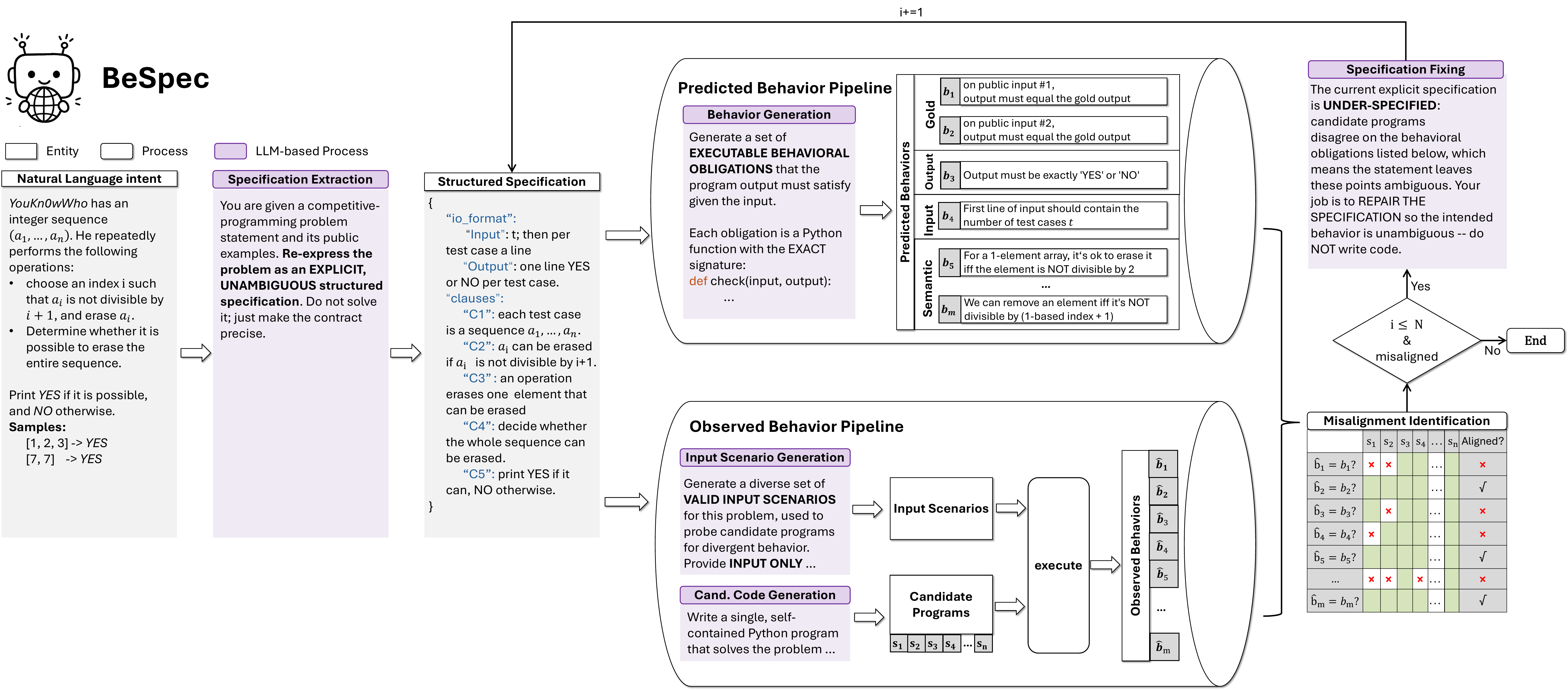}
  \caption{Overview of \sys, illustrated with the running example.}
  \label{fig:overview}
\end{figure*}

\sys treats each programming problem as an individual specification alignment task. Given the natural-language \textit{intent}, it first extracts a \emph{structured specification} (\Cref{sec:method:spec}). It then builds two pipelines for the same task (\Cref{fig:overview}). The \emph{Predicted Behavior Pipeline} describes what the intended correct program should satisfy (\Cref{sec:method:predicted}). The \emph{Observed Behavior Pipeline} captures how generated candidate programs behave on generated \emph{input scenarios} (\Cref{sec:method:real}). Subsequently, \emph{misalignment identification} compares these two views, validates mismatched behavioral expectations, and diagnoses confirmed violations as specification defects or implementation errors (\Cref{sec:method:align}). Only specification defects trigger \emph{specification fixing}, which revises the implicated clauses, regenerates candidates, and repeats the check (\Cref{sec:method:fix}). Violating candidates are penalized during \emph{behavior-grounded selection} in either case. When no specification repair is needed, or when the repair budget is exhausted, \sys selects the candidate best supported by public examples and retained behaviors (\Cref{sec:method:select}).
We use the erase-game problem from \Cref{sec:motivation} as a running example, where $[1,3]\Rightarrow$ \texttt{YES} depends on reindexing after deletion.

\subsection{Specification Extraction}
\label{sec:method:spec}

\sys first prompts the LLM to re-express the intent as an explicit \textit{structured specification} with two parts. The first gives the exact input and output layout, line by line. The second is a numbered list of \emph{clauses}. Each clause states one atomic requirement and is assigned one of seven component types: \emph{input object} (the input objects, their structure, and size), \emph{term} (a notion the statement introduces or relies on), \emph{condition} (a predicate that determines whether something holds), \emph{process} (an operation's effect on the state or its stopping condition), \emph{objective} (what is to be computed and over which set), \emph{output} (representation, special values, and order of the output), and \emph{constraint} (value ranges and input guarantees). These clauses are self-contained and directly affect output correctness.   Each clause must not solve the problem or describe an algorithm, and must be consistent with the public examples.

In the erase-game running example (\Cref{fig:overview}), the initial specification includes C1 [input object] ``each test case is a sequence $a_1,\ldots,a_n$'', C2 [condition] ``$a_i$ can be erased if $a_i$ is not divisible by $i+1$'', C3 [process] ``an operation erases one element that can be erased'', and C4 [objective] ``decide whether the whole sequence can be erased''. None of the clauses states whether the remaining elements are reindexed after a deletion. This omission leaves the ambiguity described in \Cref{sec:motivation} unresolved until later stages resolve it.

\subsection{The Predicted Behavior Pipeline}
\label{sec:method:predicted}

This pipeline starts with the \textit{structured specification} and asks the LLM to predict a set of $m$ intended behaviors, $B=\{b_1,b_2,\ldots,b_m\}$, each linked to its supporting clause. Clauses may still be ambiguous or incomplete, as the missing reindexing rule illustrates. \sys predicts and checks behaviors to expose these gaps and refine the specification. These behaviors model how the gold solution (correct program) should behave, without asking the LLM to generate it directly. This is easier than solving the full problem because a single behavior describes only an output property, an input property, or a semantic consequence of the rules. Beyond being easier to predict, these behaviors provide more detailed signals than test-level validation for specification fixing (\Cref{sec:method:fix}), because each behavior's clause link helps localize the requirement that may need repair.

Specifically, \sys predicts  four types of behaviors for the gold solution:
\begin{compactitem}
  \item \emph{Gold behaviors} are optional. They come from samples in the intent or existing tests. If such samples are provided, their exact input-output pairs must be satisfied. For instance, the erase-game example includes $[1,2,3]\Rightarrow$ \texttt{YES}.
  \item \emph{Output behaviors} state properties of the claimed output, including syntax, format, and invariants. For example, each erase-game output line must be exactly \texttt{YES} or \texttt{NO}.
  \item \emph{Input behaviors} state input-side conditions under which a behavior applies, such as boundary cases, special structures, or ambiguity-triggering scenarios. For example, in the erase-game task, the first line of the input should contain the number of test cases $t$.
  \item \emph{Semantic behaviors} state what the task rules imply for the output on a class of inputs. Some semantic behaviors cover edge cases. For example, in the erase-game task, the program should output \texttt{YES} exactly when $x$ is not divisible by $2$. Other semantic behaviors cover general rules. For example, a sequence with no deletable element should output \texttt{NO}, while a sequence that becomes erasable after earlier deletions should output \texttt{YES}.
\end{compactitem}
These four types describe what a behavior checks. A behavior can take either of two checking forms: a \emph{single behavior} checks a property of one test case's answer, while a \emph{metamorphic behavior} checks a relation between the answers to an original test case and a transformed one. Both forms are implemented as executable Python checks.
Before use, predicted behaviors that contradict public examples are revised by the LLM.

\subsection{The Observed Behavior Pipeline}
\label{sec:method:real}
In parallel, \sys executes the \textit{Observed Behavior Pipeline}, which captures how candidate programs
\emph{actually} behave when executed. This pipeline has three components: \textit{Input Scenario Generation}, \textit{Candidate Code Generation}, and \textit{Behavior Observation}.

\noindent\emph{Input Scenario Generation.} \sys generates a set of valid probe input scenarios $X=\{x_1,x_2,\ldots,x_k\}$ for observing the behaviors defined by the Predicted Behavior Pipeline. These probes provide shared scenarios on which candidate programs can be compared at the behavior level. They exercise all four behavior types defined above. By running every candidate on the same set of probes, \sys can observe where candidates agree, disagree, crash, or violate predicted behavior.

\noindent\emph{Candidate Code Generation.} \sys generates a diverse pool of candidate programs $S=\{s_1,s_2,\ldots,s_n\}$ from the structured specification. The candidates are intended to exercise the specification under the probe scenarios. Duplicate programs are removed.

\noindent\emph{Behavior Observation.} For each candidate $s_j\in S$ and each probe $x_\ell\in X$, \sys executes $s_j(x_\ell)$ and records the resulting output $y_{j,\ell}$. The execution records $Y_j=\{(x_\ell,y_{j,\ell})\}_{\ell=1}^k$ provide the evidence used to form the observed behavior vector $\hat{B}_j$ in the next step. For example, on the erase-game probe $x_\ell=[1,3]$, we observe that one candidate may produce $y_{j,\ell}=\texttt{YES}$ while another produces $\texttt{NO}$.

\subsection{Misalignment Identification}
\label{sec:method:align}

\sys compares the predicted behaviors $B=\{b_1,\ldots,b_m\}$ with the observed behavior vector $\hat{B}_j$ of each candidate $s_j$. The behavior-comparison matrix in \Cref{fig:overview} records whether each candidate matches each prediction on applicable scenarios. A behavior is \emph{aligned} if all candidates match the prediction. Otherwise, it is flagged for diagnosis. A mismatch alone does not implicate the specification because the prediction may be wrong or a candidate may be buggy.

Based on the observed outputs, \sys also asks the LLM to add behaviors for candidate disagreements not covered by existing behaviors and for public examples that many candidates fail.

\noindent\emph{Validating disputed behaviors.} \sys first determines whether each mismatched behavior is required by the original problem statement. It first consults public examples that resolve the dispute. If these are inconclusive, it uses agreement among candidates that reproduce every public example, provided such examples are available. If neither source resolves the dispute, it uses an independent LLM-as-judge. The judge sees only the original problem statement and the two competing behaviors, not the current specification or candidate code. A behavior judged not required is discarded and cannot trigger specification repair.

\noindent\emph{Diagnosing confirmed violations.} Confirmed behavior violations and failed public examples are then passed together to a single diagnosis step. This step traces each violation either to an unclear, missing, or incorrect specification clause, or to an implementation error that specification wording would not prevent. Only specification defects trigger specification fixing (\Cref{sec:method:fix}). In both cases, violating candidates are penalized at selection (\Cref{sec:method:select}). Thus, failing a public example establishes a candidate error but does not by itself establish its cause.

In the running example, candidates may answer \texttt{YES} or \texttt{NO} on $[1,3]$. After validating the reindexing requirement, diagnosis identifies the missing clause, and specification fixing makes it explicit. If the clause were already clear and the failure were an implementation error, no specification repair would follow.
\subsection{Specification Fixing}
\label{sec:method:fix}

Only violations diagnosed as specification defects are used for repair. Given the implicated clauses and triggering scenarios, \sys asks the LLM to repair the specification rather than the code. It rewrites unclear or incorrect clauses and adds new typed clauses for missing requirements, leaving unrelated clauses unchanged. Ambiguity resolutions are thus stated directly in the affected clauses. Fresh candidates are regenerated from the revised specification, and the next iteration repeats behavior comparison and diagnosis.
For the erase game, \sys{} adds a \emph{process} clause: ``after each deletion, the remaining sequence is reindexed from the front'' and regenerates candidates that now agree on $[1,3]$. \sys performs up to $N$ iterations and stops early when no violation calls for specification repair or when a repair fails to add useful new candidates.

\subsection{Behavior-Grounded Selection}
\label{sec:method:select}

On termination, \sys returns the resulting specification and one selected program. Following the intuition of CodeT~\cite{chen2023codet} and MPSC~\cite{huang2023mpsc}, \sys selects the candidate best supported by public examples and behavior checks. Specifically, it first removes candidates that fail any public sample when samples are available. Among the remaining candidates, it ranks those that satisfy more retained behaviors higher. Discarded predictions contribute no score. Thus, confirmed violations lower a candidate's rank, whether the diagnosis attributes them to the specification or the implementation. Fewer crashes or missing outputs are preferred next, and any remaining ties are broken randomly. 

\section{Experimental Design}
\label{sec:design}

\subsection{Research Questions}
\label{sec:design:rqs}

We structure the evaluation around three research questions:

\begin{compactitem}
  \item \textbf{RQ1: Effectiveness and efficiency.} How effective is \sys at improving specification alignment, and what computational cost does it require?
  Here, we assess the downstream benefit of specification alignment through program correctness, rather than measuring specification quality directly. We compare \sys with nine baselines across three LLMs and six evaluation settings, and report token and time overheads.
  \item \textbf{RQ2: Failure mode analysis.} Which problems remain hard after alignment, and why does \sys still fail on them? We first exclude problems that every method solves. We then classify the remaining failures of \sys and the strongest baselines to determine whether they stem from weak specifications or other causes, such as adopting the wrong algorithms.
  \item \textbf{RQ3: Ablation study.} What is the contribution of major components of \sys? We first examine the role of behavior prediction by comparing \sys with a variant that generates a structured specification from the intent, then generates a solution from that specification, without using behaviors. We then study the effect of candidate pool size by comparing variants with different pool sizes. Due to budget constraints, we conduct the ablation study across all datasets but only with Qwen3-Coder.
\end{compactitem}

\subsection{Datasets}
\label{sec:design:data}
We evaluate on four programming benchmarks that are standard in the
specification alignment  literature: CodeContests
\cite{li2022competition}, xCodeEval~\cite{khan2024xcodeeval}, and
APPS~\cite{hendrycks2021apps}, which together span a wide range of
difficulty and problem styles, and LiveCodeBench
\cite{jain2025livecodebench}, a continuously updated benchmark of recent contest problems. We deliberately exclude basic function completion benchmarks such as HumanEval and MBPP, on which strong models already exceed $90\%$ \pat{} with plain prompting and leave little room to study specification ambiguity. Following prior specification alignment work~\cite{specine,tian2023mufix}, we instead use harder, competition-level problems whose lengthy natural-language statements are where ambiguity actually arises.

\noindent\emph{CodeContests}~\cite{li2022competition} is a programming
benchmark curated by Google DeepMind from Codeforces and similar contest sites, comprising $13{,}328$ training, $117$ validation, and $165$ test problems. Following prior work~\cite{specine}, we use all $165$ test problems. Each problem contains two held-out private test fields: the original \emph{raw} field (which we report as \emph{CC-Raw}) and an extended field that adds roughly $190$ additional generated
tests per problem for stricter evaluation (reported as \emph{CodeContests}).

\noindent\emph{APPS}~\cite{hendrycks2021apps} collects problems from open competitive programming platforms (e.g., Codeforces, Kattis, LeetCode) across three difficulty tiers (introductory, interview, competition), with $5{,}000$ training and $5{,}000$ test problems. To balance evaluation cost and generalizability, we sample $300$ test problems stratified according to the official difficulty distribution, following prior work~\cite{specine,tian2023mufix}. We additionally rescore the same problems under \emph{APPS-Eval}~\cite{specine}, an extended test suite that augments each problem with over $100$ additional private tests.

\noindent\emph{xCodeEval}~\cite{khan2024xcodeeval} is a large, execution-based, multilingual, multitask benchmark comprising roughly $7{,}500$ competition-level problems drawn from Codeforces. In line with our APPS protocol, we sample $300$ code-generation problems stratified by the difficulty frequency distribution.

\noindent\emph{LiveCodeBench}~\cite{jain2025livecodebench} continuously collects new contest problems from platforms such as LeetCode, AtCoder, and Codeforces, with each problem tagged by release date. We use the LiveCodeBench split containing $112$ recent problems released on or after 2025-01-01, which helps mitigate data leakage.

In total, we report \emph{six} evaluation settings: the four benchmarks plus the two extended test suites (CodeContests, APPS-Eval). All methods are evaluated on the held-out private test sets for each problem. We use the same problem instances and metrics for all methods, ensuring a fair comparison.

\subsection{Metrics}
\label{sec:design:metrics}

Following prior work~\cite{specine,tian2023mufix}, we report two effectiveness metrics. \pat{} is the fraction of problems whose selected
program passes \emph{all} private tests, which is the strict, deployment-relevant metric. \apr{} (average pass rate) is the average fraction of private tests passed, a finer-grained signal that credits partial progress and is less sensitive to a single failing edge case. 

We further report two efficiency metrics~\cite{specine}: token overhead, defined as the total number of prompt and completion tokens consumed per problem, and time overhead, defined as the end-to-end wall-clock time per problem. Smaller values indicate better efficiency. Because CodeContests and APPS reuse the same generated programs under their alternative scoring fields (CC-Raw, APPS-Eval), average efficiency is aggregated over the four distinct benchmarks and reported per method alongside effectiveness
in \Cref{tab:accuracy}.

\subsection{Baselines}
\label{sec:design:baselines}

In this work, we compare \sys against nine state-of-the-art baselines covering the main families of LLM code generation. We select baselines with available replication packages. ReaCoder~\cite{reacoder} and SpecAlign~\cite{sepcalign} are not included because their replication packages were not available at the time of writing. Specifically, our baselines are as follows. \textit{Zero-shot prompting} measures the base model's performance.
\textit{SCoT}~\cite{li2023scot} asks the model to produce a structured chain-of-thought plan before coding. \textit{Self-Repair}~\cite{olausson2023selfrepair} iteratively revises code using execution feedback. \textit{Self-Collaboration}~\cite{dong2024selfcollab} and \textit{AgentCoder}~\cite{huang2023agentcoder} use multi-agent workflows with roles such as analyst, coder, and tester. \textit{PairCoder}~\cite{zhang2024paircoder} models pair programming with a navigator and a driver agent. \textit{$\mu$FiX}~\cite{tian2023mufix} uses test case analysis and feedback prompting to improve the model's understanding of the specification.

Our most direct comparisons are Specine~\cite{specine} and
SpecFix~\cite{specfix}, two specification alignment methods that share our motivation. Specine lifts the specification perceived by the LLM from the generated code and aligns it with the input using requirements engineering rules. SpecFix repairs the program distribution induced by an ambiguous requirement and maps the change back to the requirement through contrastive specification inference. These methods represent
the current state-of-the-art in specification alignment. Unlike \sys, however, they rely on test-level feedback: they run candidate code against tests and revise the intent based on the resulting failures. \sys instead works at the behavior level, decomposing the intent into behaviors that are easier to predict, check, localize, and refine into a better specification.

\subsection{Implementation and Setup}
\label{sec:design:impl}

We implement \sys and all baselines on three advanced LLMs to evaluate whether the method generalizes across model families:
\texttt{qwen3-coder-30b}, an open-weight coding-specialized LLM;
\texttt{glm-4.7-flash}, an open-weight reasoning-oriented LLM; and
\texttt{gpt-5-mini}, a proprietary general-purpose LLM. Following prior practice~\cite{specine}, we set the temperature to $0.8$ for \texttt{qwen3-coder-30b} and \texttt{glm-4.7-flash}. For \texttt{gpt-5-mini}, we use the default temperature because custom temperature settings are not supported. We set the maximum number of iterations $N$ to $3$ for both \sys and the baselines. We set the pool size for candidate solutions to be 20 for SpecFix and \sys, as suggested by the former \cite{specfix}. We run the experimental pipelines and execute candidate programs on a Dell Precision 7960 Tower workstation with an Intel Xeon w9-3495X processor. We run experiments concurrently across multiple problems, with all LLM calls made through the OpenRouter API~\cite{openrouter}. Our replication package is available on GitHub~\footnote{\url{https://github.com/qiqiguana/bespec}}.

\section{Results}
\label{sec:results}

\subsection{RQ1: Overall Performance}
\label{sec:results:rq1}
\subsubsection{Effectiveness}
\begin{table*}[t]
  \centering
  \caption{Effectiveness and efficiency of \sys and nine baselines across three
  LLMs and six evaluation settings. \textbf{CC-Raw} and \textbf{CodeContests}
  denote the raw and extended CodeContests private-test fields, respectively.
  \textbf{APPS-Eval} denotes the extended APPS evaluation field, and
  \textbf{LCB} denotes LiveCodeBench. \pat{} ($\uparrow$) is the
  fraction of problems passing all private tests; \apr{} ($\uparrow$) is the
  average private-test pass rate. \textbf{Tok} ($\downarrow$) is measured in
  $10^3$ tokens and \textbf{Time} ($\downarrow$) in seconds. \textbf{Avg.}
  reports the mean \pat{} and \apr{} across the six evaluation settings.
  $\Delta_{\mathrm{rel}}$ reports the relative change of \sys over the best
  baseline in each effectiveness column and over SpecFix for token and time costs.}
  \label{tab:accuracy}
  \scriptsize
  \setlength{\tabcolsep}{1.25pt}
  \begin{tabular}{ll cc cc cc cc cc cc cc cc}
    \toprule
    & & \multicolumn{14}{c}{Effectiveness} & \multicolumn{2}{c}{\multirow{2}{*}{Efficiency}} \\
    \cmidrule(lr){3-16}
    & & \multicolumn{2}{c}{CC-Raw} & \multicolumn{2}{c}{CodeContests}
    & \multicolumn{2}{c}{xCodeEval} & \multicolumn{2}{c}{APPS}
    & \multicolumn{2}{c}{APPS-Eval} & \multicolumn{2}{c}{LCB}
    & \multicolumn{2}{c}{Avg.} \\
    \cmidrule(lr){3-4}\cmidrule(lr){5-6}\cmidrule(lr){7-8}\cmidrule(lr){9-10}\cmidrule(lr){11-12}\cmidrule(lr){13-14}\cmidrule(lr){15-16}\cmidrule(lr){17-18}
    LLM & Method & \pat{} & \apr{} & \pat{} & \apr{} & \pat{} & \apr{} & \pat{} & \apr{}
           & \pat{} & \apr{} & \pat{} & \apr{} & \pat{} & \apr{} & Tok & Time \\
    \midrule
    \multirow{11}{*}{\texttt{qwen3-coder-30b}} & Zero-shot & 5.5 & 8.0 & 5.5 & 8.1 & 28.0 & 33.4 & 35.0 & 39.7 & 13.7 & 34.4 & 27.7 & 31.0 & 19.2 & 25.8 & 1.4 & 16 \\
     & SCoT & 21.8 & 30.3 & 17.0 & 28.0 & 29.7 & 39.6 & 43.7 & 55.4 & 18.0 & 47.9 & 35.7 & 46.6 & 27.7 & 41.3 & 3.2 & 28 \\
     & Self-Repair & 20.6 & 28.4 & 17.0 & 27.9 & 35.3 & 45.5 & 52.3 & 61.9 & 21.0 & 53.6 & 37.5 & 44.6 & 30.6 & 43.6 & 10.7 & 98 \\
     & Self-Collaboration & 24.2 & 33.5 & 20.6 & 32.0 & 39.3 & 51.0 & 52.3 & 62.7 & 21.0 & 54.9 & 39.3 & 48.2 & 32.8 & 47.1 & 8.4 & 56 \\
     & AgentCoder & 13.3 & 18.8 & 13.3 & 18.0 & 37.3 & 43.4 & 48.3 & 52.5 & 18.7 & 46.8 & 38.4 & 44.5 & 28.2 & 37.3 & 10.3 & 106 \\
     & $\mu$FiX & 22.4 & 31.6 & 17.6 & 31.5 & 37.7 & 49.7 & 51.3 & 64.2 & 19.7 & 55.4 & 36.6 & 48.7 & 30.9 & 46.9 & 8.4 & 60 \\
     & PairCoder & 24.8 & 36.3 & 21.8 & 36.1 & 42.3 & 53.5 & 58.7 & 69.2 & 24.3 & 58.4 & 35.7 & 48.5 & 34.6 & 50.3 & 23.4 & 162 \\
     & SpecFix & 24.2 & 27.7 & 20.0 & 27.6 & 36.3 & 41.6 & 52.0 & 56.5 & 20.7 & 49.8 & 37.5 & 42.0 & 31.8 & 40.9 & 129.1 & 316 \\
     & Specine & 27.9 & 35.1 & 23.6 & 33.8 & 41.0 & 53.3 & 58.0 & 68.0 & 23.7 & 59.0 & 39.3 & 48.8 & 35.6 & 49.7 & 25.4 & 169 \\
     & \sys & \textbf{42.4} & \textbf{56.0} & \textbf{33.9} & \textbf{54.5} & \textbf{46.3} & \textbf{60.0} & \textbf{68.0} & \textbf{80.2} & \textbf{28.0} & \textbf{68.2} & \textbf{45.5} & \textbf{60.2} & \textbf{44.0} & \textbf{63.2} & 60.3 & 364 \\
    \rowcolor{gray!10}
     & $\Delta_{\mathrm{rel}}$ & +52.0\% & +54.3\% & +43.6\% & +51.0\% & +9.5\% & +12.1\% & +15.8\% & +15.9\% & +15.2\% & +15.6\% & +15.8\% & +23.4\% & +23.6\% & +25.6\% & -53.3\% & +15.2\% \\
    \midrule
    \multirow{11}{*}{\texttt{glm-4.7-flash}} & Zero-shot & 8.5 & 14.1 & 6.7 & 14.3 & 25.3 & 33.9 & 38.3 & 48.3 & 14.3 & 42.5 & 28.6 & 36.2 & 20.3 & 31.6 & 3.5 & 41 \\
     & SCoT & 12.7 & 19.8 & 9.7 & 19.7 & 26.3 & 37.1 & 34.7 & 50.9 & 13.7 & 43.3 & 29.5 & 39.0 & 21.1 & 35.0 & 8.1 & 92 \\
     & Self-Repair & 16.4 & 26.0 & 13.3 & 24.1 & 37.7 & 49.8 & 47.7 & 61.7 & 18.7 & 53.1 & 32.1 & 50.2 & 27.7 & 44.1 & 25.0 & 290 \\
     & Self-Collaboration & 21.2 & 29.6 & 17.0 & 30.2 & 35.0 & 49.5 & 49.7 & 63.1 & 20.3 & 54.4 & 32.1 & 50.2 & 29.2 & 46.2 & 31.9 & 355 \\
     & AgentCoder & 21.8 & 30.8 & 17.0 & 28.7 & 43.0 & 55.1 & 50.3 & 61.1 & 18.7 & 51.2 & 38.4 & 49.9 & 31.5 & 46.1 & 29.6 & 440 \\
     & $\mu$FiX & 20.6 & 29.3 & 18.2 & 27.7 & 32.0 & 47.2 & 49.3 & 59.5 & 19.3 & 51.7 & 33.0 & 44.6 & 28.7 & 43.3 & 21.8 & 249 \\
     & PairCoder & 23.6 & 34.7 & 17.6 & 32.7 & 38.7 & 50.1 & 53.0 & 68.3 & 21.3 & 56.4 & 39.3 & 55.2 & 32.2 & 49.6 & 67.6 & 993 \\
     & SpecFix & 21.8 & 26.7 & 18.2 & 27.9 & 36.3 & 45.8 & 51.3 & 58.8 & 19.3 & 50.4 & 34.8 & 40.7 & 30.3 & 41.7 & 292.7 & 1614 \\
     & Specine & 29.1 & 40.4 & 25.5 & 38.1 & 40.3 & 53.5 & 53.0 & 65.4 & 20.7 & 53.9 & 40.2 & 56.1 & 34.8 & 51.2 & 61.9 & 833 \\
     & \sys & \textbf{35.8} & \textbf{54.0} & \textbf{27.9} & \textbf{50.9} & \textbf{45.7} & \textbf{64.3} & \textbf{63.3} & \textbf{80.4} & \textbf{28.0} & \textbf{67.0} & \textbf{47.3} & \textbf{58.5} & \textbf{41.3} & \textbf{62.5} & 122.1 & 1132 \\
    \rowcolor{gray!10}
     & $\Delta_{\mathrm{rel}}$ & +23.0\% & +33.7\% & +9.4\% & +33.6\% & +6.3\% & +16.7\% & +19.4\% & +17.7\% & +31.5\% & +18.8\% & +17.7\% & +4.3\% & +18.7\% & +22.1\% & -58.3\% & -29.9\% \\
    \midrule
    \multirow{11}{*}{\texttt{gpt-5-mini}} & Zero-shot & 47.9 & 51.4 & 46.1 & 49.6 & 65.7 & 71.3 & 72.3 & 78.0 & 29.3 & 64.0 & 42.9 & 50.6 & 50.7 & 60.8 & 3.4 & 52 \\
     & SCoT & 43.0 & 49.7 & 39.4 & 47.9 & 64.0 & 71.0 & 74.7 & 80.8 & 31.0 & 64.6 & 42.0 & 49.1 & 49.0 & 60.5 & 6.9 & 73 \\
     & Self-Repair & 57.0 & 62.8 & 50.9 & 58.6 & 70.7 & 76.2 & 78.7 & 84.5 & 33.3 & 68.2 & 44.6 & 51.8 & 55.9 & 67.0 & 14.9 & 220 \\
     & Self-Collaboration & 49.7 & 56.1 & 43.6 & 51.1 & 70.3 & 75.2 & 80.3 & 85.6 & 34.0 & 69.8 & 47.3 & 58.3 & 54.2 & 66.0 & 18.3 & 227 \\
     & AgentCoder & 69.1 & 76.2 & 61.8 & 71.7 & 70.0 & 74.4 & 85.7 & 90.4 & 35.7 & 73.0 & 55.4 & 64.5 & 62.9 & 75.0 & 20.3 & 336 \\
     & $\mu$FiX & 55.2 & 61.6 & 50.3 & 59.2 & 68.7 & 74.9 & 82.3 & 87.2 & 34.7 & 70.4 & 46.4 & 54.4 & 56.3 & 68.0 & 12.8 & 156 \\
     & PairCoder & 72.7 & 79.5 & 63.6 & 72.5 & 71.7 & 78.0 & 84.3 & 90.3 & 33.7 & 71.5 & 61.6 & 71.0 & 64.6 & 77.1 & 31.0 & 382 \\
     & SpecFix & 66.1 & 70.0 & 60.6 & 67.2 & 72.0 & 77.9 & 81.3 & 86.3 & 34.3 & 69.4 & 56.2 & 63.2 & 61.8 & 72.3 & 225.5 & 1104 \\
     & Specine & 70.9 & 78.3 & 63.0 & 72.2 & 69.7 & 77.4 & 86.0 & 91.4 & 36.3 & 73.0 & 54.5 & 66.2 & 63.4 & 76.4 & 33.5 & 456 \\
     & \sys & \textbf{78.8} & \textbf{87.5} & \textbf{72.1} & \textbf{80.3} & \textbf{76.3} & \textbf{82.3} & \textbf{88.0} & \textbf{93.1} & \textbf{37.3} & \textbf{75.5} & \textbf{71.4} & \textbf{85.4} & \textbf{70.6} & \textbf{84.0} & 104.7 & 853 \\
    \rowcolor{gray!10}
     & $\Delta_{\mathrm{rel}}$ & +8.4\% & +10.1\% & +13.4\% & +10.8\% & +6.0\% & +5.5\% & +2.3\% & +1.9\% & +2.8\% & +3.4\% & +15.9\% & +20.3\% & +9.3\% & +8.9\% & -53.6\% & -22.7\% \\
    \bottomrule
  \end{tabular}
\end{table*}

\Cref{tab:accuracy} reports \pat{} and \apr{} for all methods across the three
LLMs and six evaluation settings. We analyze the results from two complementary
perspectives: whether the gains hold \emph{across LLMs}, and whether
they hold \emph{across benchmarks}.

\textbf{Across LLMs.} The \textit{Avg.} columns summarize performance over all six
evaluation settings and show that \sys achieves the highest mean \pat{} and
\apr{} for every LLM. As expected, all methods improve when moving from the
smaller open-weight models (\texttt{qwen3-coder-30b} and
\texttt{glm-4.7-flash}) to the stronger proprietary model
(\texttt{gpt-5-mini}), reflecting the effect of base model capability. However,
\sys consistently outperforms the baselines across LLMs. Relative to
the strongest baseline, \sys improves mean \pat{} and \apr{} by $23.6\%$ and
$25.6\%$ on \texttt{qwen3-coder-30b}, by $18.7\%$ and $22.1\%$ on
\texttt{glm-4.7-flash}, and by $9.3\%$ and $8.9\%$ on \texttt{gpt-5-mini}. This
indicates that \sys generalizes across a coding-specialized LLM, a
reasoning-oriented LLM, and a strong proprietary LLM. The relative gains are largest on the open-weight Qwen and GLM backbones, suggesting that \sys can partially
compensate for weaker base model capability through explicit specification
alignment. This makes \sys especially useful when only smaller LLMs are
accessible due to resource-related or financial constraints.

\textbf{Across benchmarks.}
\sys achieves the state of the art \pat{} and \apr{} on all benchmarks. On \texttt{qwen3-coder-30b}, \sys improves \pat{} by $52.0\%$ relative
on CC-Raw ($27.9\%\rightarrow42.4\%$) and by $15.8\%$ on APPS
($58.7\%\rightarrow68.0\%$). On \texttt{glm-4.7-flash}, it improves APPS
\pat{} by $19.4\%$ ($53.0\%\rightarrow63.3\%$). On \texttt{gpt-5-mini}, it
improves CC-Raw \pat{} by $8.4\%$ ($72.7\%\rightarrow78.8\%$). On xCodeEval,
\sys improves \pat{} over the best baseline by $9.5\%$, $6.0\%$, and $6.3\%$
relative on \texttt{qwen3-coder-30b}, \texttt{gpt-5-mini}, and
\texttt{glm-4.7-flash}, respectively. The corresponding \apr{} gains are
$12.1\%$, $5.5\%$, and $16.7\%$, showing that \sys's advantage generalizes across
benchmarks.

The extended private test suites are stricter than their raw counterparts.
CodeContests extends CC-Raw, and APPS-Eval extends APPS. Both score the same
generated programs with more complete private test fields. As a result, scores
drop for most methods when moving from CC-Raw to CodeContests and from APPS to
APPS-Eval, showing that the extended suites expose failures missed by the raw
test fields. Despite this stricter evaluation, \sys remains the best method in every cell on both metrics. On CodeContests, it improves \pat{} over the strongest baseline by
$43.6\%$, $9.4\%$, and $13.4\%$ relative on \texttt{qwen3-coder-30b},
\texttt{glm-4.7-flash}, and \texttt{gpt-5-mini}, respectively. On APPS-Eval, the
corresponding gains are $15.2\%$, $31.5\%$, and $2.8\%$. This stability across raw and extended test suites suggests that \sys improves the underlying program behavior rather than overfitting to specific evaluation test suites.

Finally, we observe the same pattern on LiveCodeBench, the
most recent benchmark. \sys improves \pat{} over the best baseline by
$15.8\%$, $17.7\%$, and $15.9\%$ relative on \texttt{qwen3-coder-30b},
\texttt{glm-4.7-flash}, and \texttt{gpt-5-mini}, respectively.
Because these problems are more recent, this result suggests that the gains are
less likely to come from memorization and more likely to reflect better
specification alignment.

\subsubsection{Efficiency} The two rightmost columns of \Cref{tab:accuracy} report the mean token and time cost per problem. Most baselines are relatively inexpensive. Excluding SpecFix and Specine, they use at most $67.6$k tokens and $993$ seconds
per problem on average. SpecFix is substantially more token-intensive than the
other methods, using $129.1$k to $292.7$k tokens per problem, and Specine also
incurs a higher cost than lightweight baselines. Compared with SpecFix, \sys is
both more effective and less token-intensive. For example, on
\texttt{gpt-5-mini}, \sys uses $104.7$k tokens per problem compared with
$225.5$k for SpecFix. Compared with Specine, \sys is more costly because it
executes a pool of candidate programs against probe scenarios. However, the
absolute cost remains practical. Across the three LLMs, \sys uses $60.3$k to
$122.1$k tokens and takes $364$ to $1132$ seconds per problem on average, which
is acceptable for program generation workloads where accuracy is the
primary goal.

\rqconclusion{RQ1}{\sys consistently improves mean \pat{} and \apr{} across all
LLMs and benchmarks, with relative gains of $9.3$ to $23.6\%$ and $8.9$ to $25.6\%$ over the strongest baselines, while incurring an acceptable cost of
$60.3$k to $122.1$k tokens and $364$ to $1132$ seconds per problem.}

\subsection{RQ2: What Remains After Alignment?}
\label{sec:results:rq2}

RQ1 shows that \sys improves performance, but it does not explain
why the remaining problems are unsolved. We therefore analyze residual failures in
two steps. First, we identify the problem--model pairs that are not solved by every method
(\Cref{sec:results:rq2:hard}). Second, we analyse the failure modes of \sys and two specification alignment baselines (Specine and SpecFix) (\Cref{sec:results:rq2:tax}).
Both analyses use CodeContests, APPS, xCodeEval, and LCB
consistently across all LLMs, with ``solved'' defined as
passing all private tests.

\subsubsection{Which problems remain hard?}
\label{sec:results:rq2:hard}

\begin{figure}[t]
  \centering
  \includegraphics[width=\columnwidth]{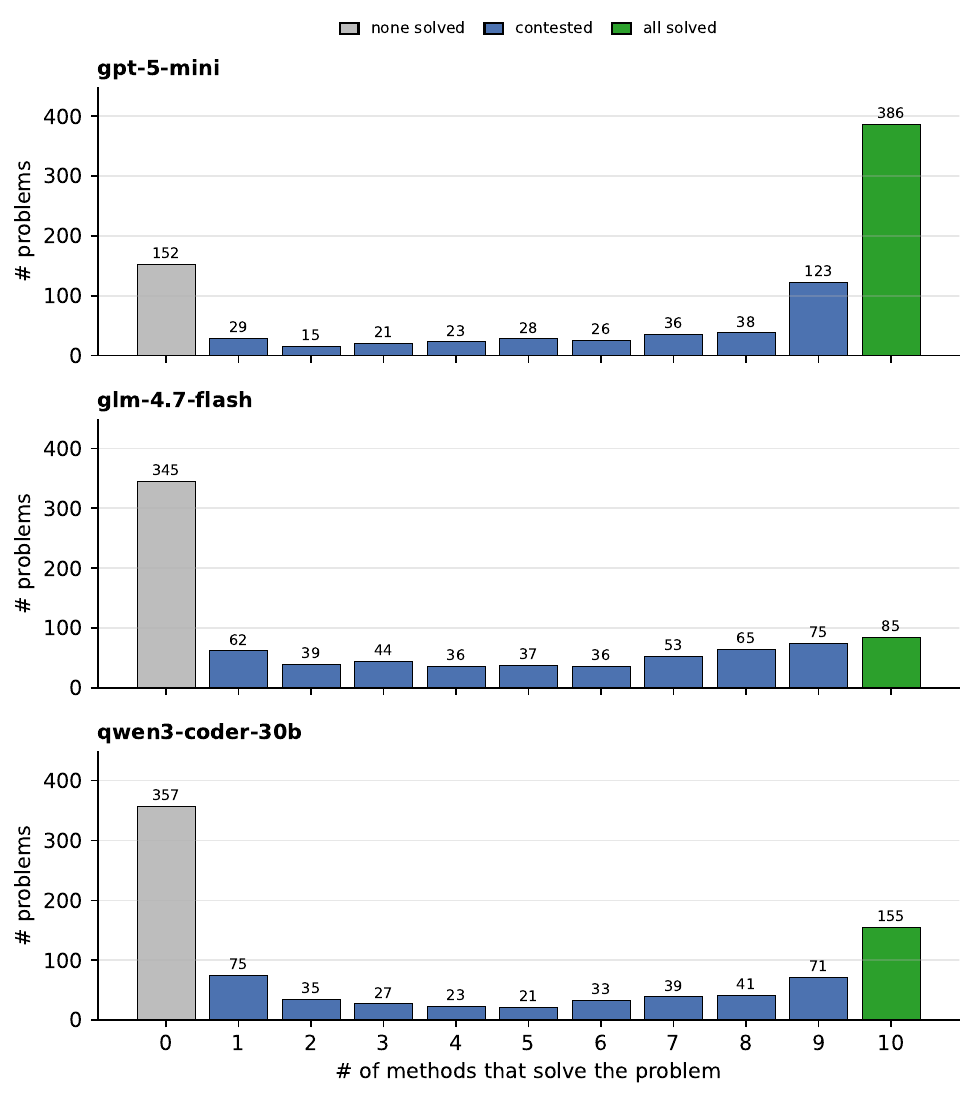}
  \caption{Distribution of problem--model pairs by the number of methods that solve each pair.}
  \label{fig:solved}
\end{figure}

For each problem--model pair, we count how many of the ten methods (\sys plus the nine
baselines in \Cref{tab:accuracy}) solve it.
\Cref{fig:solved} shows the distribution across $2{,}631$ pairs ($877$ problems $\times$ $3$ LLMs).
Of these, $626$ pairs ($23.8\%$) are solved by all ten
methods, indicating easy instances for the corresponding LLM.
Another $854$ pairs ($32.5\%$) are solved by none of the methods, indicating
instances that remain unsolved by all evaluated methods with the corresponding LLM.
Method differences are therefore determined by the $1{,}151$ contested pairs
($k\!=\!1..9$), where some methods succeed, and others fail. We exclude the easy
set and analyze failures on the remaining hard pairs.

\subsubsection{A failure mode analysis}
\label{sec:results:rq2:tax}

\begin{figure}[t]
  \centering
  \includegraphics[width=\columnwidth]{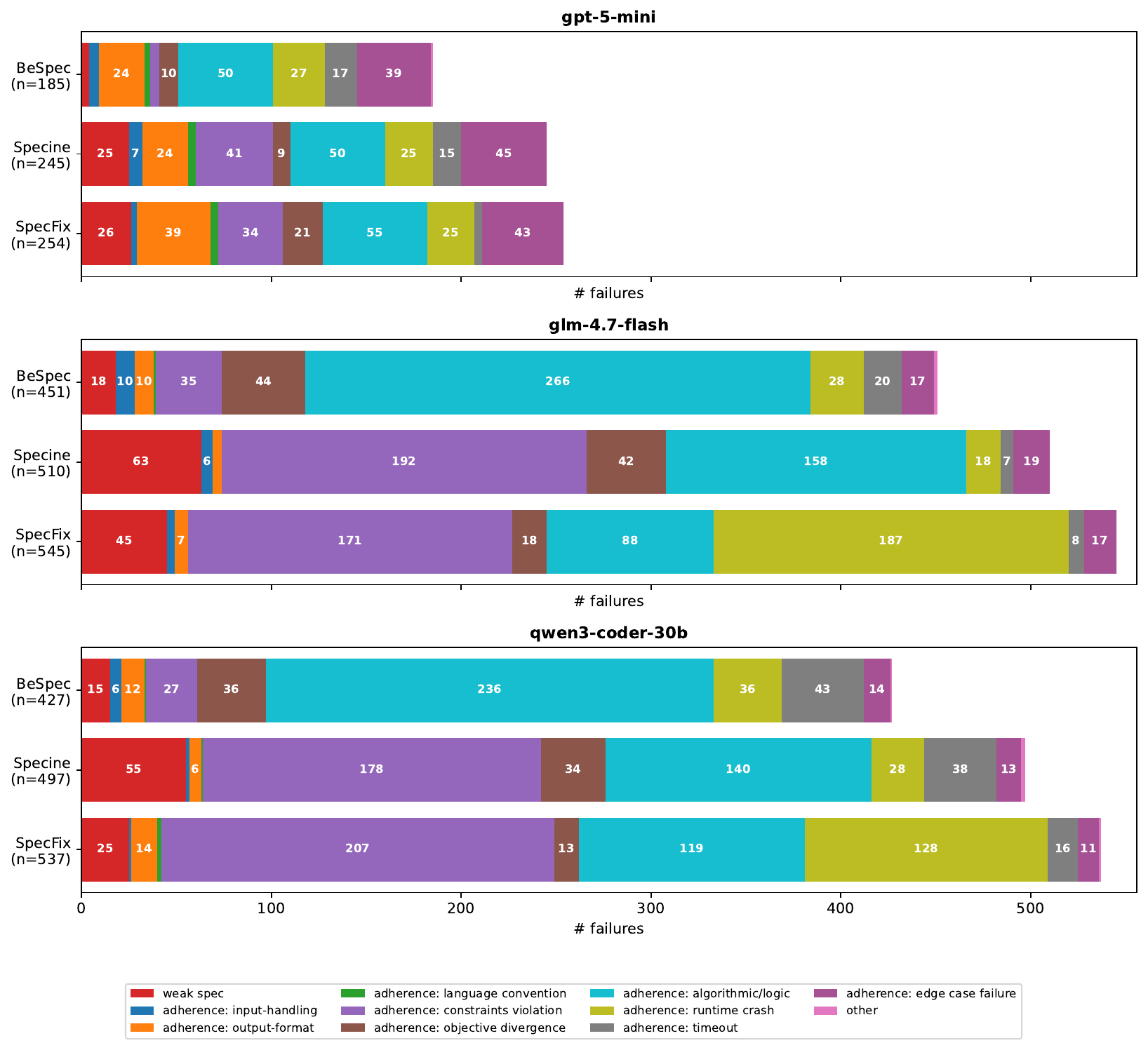}
  \caption{Failure mode analysis of \sys, Specine, and SpecFix on hard problems.}
  \label{fig:taxonomy}
\end{figure}

We analyse the hard problem--model pairs for the three specification alignment
methods only, namely \sys, Specine, and SpecFix. These three methods are among the best-performing approaches in several settings (\Cref{tab:accuracy}). The more informative question
is therefore whether the remaining failures of the three specification alignment methods are still caused by specification errors. Specifically, we analyse the failure types shown in \Cref{fig:taxonomy}:

\begin{compactitem}
  \item \emph{Weak spec.} The refined specification still omits or
  misstates something required by the problem statement, such as the input
  format, output format, a stated constraint, or the objective.
  \item \emph{Adherence.} The refined specification already states the behavior
  needed to pass the failing test, but the selected program does not follow it.
  We divide adherence failures into nine subtypes: \emph{input-handling} parses
  the input incorrectly; \emph{output-format} computes the right value but returns
  it in the wrong form; \emph{language convention} misuses programming-language
  conventions, such as integer division, modulo behavior, or array bounds;
  \emph{constraints violation} ignores a stated bound or validity condition; \emph{objective divergence}  computes the wrong target; \emph{algorithmic/logic} uses an incorrect algorithm or control logic; \emph{runtime crash} terminates with an  exception or invalid operation; \emph{timeout} exceeds the time limit; and
  \emph{edge case failure} passes visible tests but fails a hidden case.
  \item \emph{Other.} The failure does not fit weak spec or any adherence
  subtype, such as an LLM response parsing failure.
\end{compactitem}

We classify every residual failure with an LLM judge (\texttt{gpt-5-mini},
using the default temperature). The judge reads the statement, the predicted specification, the
selected program, and the failing tests, then emits exactly one class. To validate
the judge, we drew a fixed seed sample of $100$ failures, stratified across all
classes and the three specification alignment systems. One author labeled these
failures while blind to the model output. Human and judge labels agree on $80\%$
of the sample, with Cohen's $\kappa=0.78$. Under the Landis and Koch
interpretation~\cite{landis1977measurement} of Cohen's $\kappa$
statistic~\cite{cohen1960kappa}, this indicates substantial agreement. This validation
supports using the LLM judge for the full failure set, consistent with prior LLM
judge protocols~\cite{zheng2023judging}. \Cref{fig:taxonomy} shows the resulting
distribution.

The taxonomy shows that, among problems that remain hard, the specification is no
longer the main bottleneck. Across all three specification alignment methods,
\emph{weak spec} is a small minority of failures. For \sys, \textit{weak specification} failures
account for $4$ out of $185$ failures on \texttt{gpt-5-mini}, $18$ out of $451$
on \texttt{glm-4.7-flash}, and $15$ out of $427$ on
\texttt{qwen3-coder-30b}, for a total of $37$ out of $1{,}063$. The
corresponding totals are higher for the baselines: Specine has $25+63+55=143$
\textit{weak specification} failures out of $1{,}252$, and SpecFix has $26+45+25=96$ out of
$1{,}336$. This indicates that all three specification alignment methods improve
the generated specifications, and that \sys produces the fewest\textit{ weak specification} failures
among them. The remaining \sys failures are concentrated in
\emph{adherence: algorithmic/logic}, \emph{adherence: runtime crash}, and \emph{adherence: timeout}, where the program reflects the intended specification but uses an incorrect or inefficient algorithm. Thus, after intent is aligned, further performance gains require better implementations. This may come from more powerful LLMs or from code refinement that better enforces adherence to the refined specification.

\rqconclusion{RQ2}{After alignment, failures are rarely due to weak specifications. Only $37$ out of $1{,}063$ of \sys's remaining failures are classified as \emph{weak spec}, compared with $143$ out of $1{,}252$ for Specine and $96$ out of $1{,}336$ for SpecFix, while most remaining errors are adherence failures.}

\subsection{RQ3: Ablation Study}
\label{sec:results:rq3}

For the ablation study, we ask two questions. First, does the \emph{behavior} model in \sys improve performance (\Cref{sec:results:rq3:behavior})?
Second, how does the candidate pool size affect performance
(\Cref{sec:results:rq3:pool})? To save budget, we run all ablations on all
benchmarks but only with \texttt{qwen3-coder-30b}.
\subsubsection{Ablation study on the behavior model}
\label{sec:results:rq3:behavior}

To assess whether the behavior model is useful, we compare \sys with a variant that
does not use behaviors. This variant generates a structured specification from
the intent, then generates a solution from that specification. It does not
predict behaviors, check candidate programs against behaviors, or use behavior
feedback to refine the specification. \Cref{tab:ablation} compares this variant
($\mathcal{V}_{\text{w/o bh}}$) with the full system on the same problems.

\begin{table}[t]
  \centering
  \caption{Ablations on \texttt{qwen3-coder-30b}. $\mathcal{V}_{\text{w/o bh}}$ removes the behavior model, $\mathcal{V}_{n=k}$ uses pool size $k$, and \emph{Full} is \sys. \emph{Iters} denotes mean refinement iterations. \emph{CC} and \emph{xCE} denote \textit{CodeContests} and \textit{xCodeEval}.}
  \label{tab:ablation}
  \scriptsize
  \setlength{\tabcolsep}{3pt}
  \resizebox{\columnwidth}{!}{%
  \begin{tabular}{l c cc cc cc cc cc cc}
    \toprule
    & & \multicolumn{2}{c}{CC-Raw} & \multicolumn{2}{c}{CC}
    & \multicolumn{2}{c}{xCE} & \multicolumn{2}{c}{APPS}
    & \multicolumn{2}{c}{APPS-Eval} & \multicolumn{2}{c}{LCB} \\
    \cmidrule(lr){3-4}\cmidrule(lr){5-6}\cmidrule(lr){7-8}\cmidrule(lr){9-10}\cmidrule(lr){11-12}\cmidrule(lr){13-14}
    Variant & Iters & P@1 & \apr & P@1 & \apr & P@1 & \apr & P@1 & \apr
            & P@1 & \apr & P@1 & \apr \\
    \midrule
    $\mathcal{V}_{\text{w/o bh}}$ & -- & 17.0 & 27.0 & 15.2 & 26.0 & 36.3 & 45.6 & 46.3 & 57.1
                 & 20.0 & 50.8 & 33.9 & 42.6 \\
    \midrule
    $\mathcal{V}_{n=2}$  & 1.97 & 34.5 & 45.9 & 24.2 & 41.5 & 42.7 & 55.6 & 61.7 & 74.5 & 25.0 & 63.4 & 44.6 & 57.2 \\
    $\mathcal{V}_{n=4}$  & 1.84 & 38.2 & 51.1 & 29.7 & 49.8 & 45.7 & 59.0 & 64.3 & 75.7 & 28.0 & 63.1 & 46.4 & 59.1 \\
    $\mathcal{V}_{n=6}$  & 1.79 & 33.9 & 48.5 & 27.9 & 45.9 & 45.0 & 58.0 & 64.3 & 77.6 & 26.7 & 66.3 & 45.5 & 59.1 \\
    $\mathcal{V}_{n=8}$  & 1.78 & 32.1 & 46.4 & 23.0 & 44.3 & 46.7 & 60.2 & 61.3 & 74.9 & 25.7 & 63.5 & 45.5 & 60.2 \\
    $\mathcal{V}_{n=10}$ & 1.66 & 42.4 & 54.1 & 33.3 & 50.9 & 46.0 & 60.0 & 65.0 & 77.0 & 26.3 & 66.1 & 45.5 & 59.5 \\
    $\mathcal{V}_{n=12}$ & 1.66 & 38.8 & 52.1 & 31.5 & 50.0 & 49.3 & 60.9 & 64.7 & 78.7 & 28.3 & 67.8 & 43.8 & 58.7 \\
    $\mathcal{V}_{n=14}$ & 1.68 & 40.0 & 54.7 & 28.5 & 50.3 & 46.3 & 61.1 & 64.3 & 79.5 & 28.7 & 67.7 & 45.5 & 59.8 \\
    $\mathcal{V}_{n=16}$ & 1.61 & 38.2 & 50.5 & 29.1 & 49.2 & 49.3 & 62.8 & 65.7 & 79.1 & 28.0 & 67.6 & 46.4 & 60.1 \\
    $\mathcal{V}_{n=18}$ & 1.60 & 41.8 & 55.4 & 32.7 & 52.1 & 46.3 & 60.2 & 66.7 & 80.3 & 27.7 & 69.3 & 42.9 & 60.5 \\
    \midrule
    Full & 1.58 & 42.4 & 56.0 & 33.9 & 54.5 & 46.3 & 60.0 & 68.0 & 80.2 & 28.0 & 68.2 & 45.5 & 60.2 \\
    \bottomrule
  \end{tabular}}
\end{table}

We observe that the behavior model contributes substantially and consistently across all six settings. On CC, removing the behavior model reduces \pat{} by $55.2\%$
and \apr{} by $52.3\%$. On CC-Raw, the reductions are $59.9\%$ and $51.8\%$. On xCodeEval, they are $21.6\%$ and $24.0\%$.
On APPS, they are $31.9\%$ and $28.8\%$. On APPS-Eval, they are $28.6\%$
and $25.5\%$. On LiveCodeBench, they are $25.5\%$
and $29.2\%$. Thus, removing
behaviors hurts both exact correctness and partial correctness in every setting,
showing that the behavior model is a central contributor to \sys's gains.

\subsubsection{Ablation study on the pool size}
\label{sec:results:rq3:pool}

We next study the effect of candidate pool size. We vary the pool size
$n\in\{2,4,\dots,20\}$ while keeping the maximum iteration budget
fixed at $N{=}3$, as in RQ1, with the early-stopping conditions in
\Cref{sec:method:fix}. \Cref{tab:ablation} reports these pool-size variants
across all benchmarks. The Full row uses the same configuration as RQ1
($n=20$, $N=3$) and reports the same results.

We observe an overall tendency for larger pools to improve accuracy, although
results fluctuate across pool sizes, with many gains appearing by pool size $10$. On CC-Raw, increasing the
pool from $2$ to $20$ improves \pat{} from $34.5$ to $42.4$ and \apr{} from
$45.9$ to $56.0$. Similar gains appear in several other settings, while LCB remains relatively
stable. At the same time, the average number of refinement iterations generally
decreases as the pool grows, from $1.97$ at pool
$2$ to $1.58$ at pool $20$. Thus, pool size and refinement iterations trade off:
a larger pool is more likely to include a strong candidate early, while a smaller
pool relies more on specification refinement to improve the specification and
the subsequently generated code over additional repair iterations.

With the same maximum iteration budget as RQ1 (\Cref{sec:results:rq1}),
smaller pools can achieve similar results in some settings. In practice,
the pool size and iteration budget should therefore
be chosen according to the available compute constraints. When concurrency is
available, a larger pool with fewer refinement iterations can exploit parallel
execution. Otherwise, a smaller pool with more refinement iterations can reduce
peak resource usage.

\rqconclusion{RQ3}{Removing the behavior model substantially reduces both \pat{}
and \apr{}. Pool size can trade off with refinement iterations, so larger pools
with fewer iterations may be preferable when concurrency is available, while smaller
pools with more iterations may be preferable otherwise.}
\section{Discussion}
\label{sec:discussion}

\subsection{Specification Refinement versus Candidate Selection}
\label{sec:disc-sampling}

\sys refines specifications based on behavioral feedback, but it also
generates a pool of candidate programs in the process. Thus, its gains may stem more from exploring additional implementation variants than from refining the specification. To examine the contribution
of the candidate pool, we use 40 problems per LLM, with 10 from each
benchmark. For each problem, we sample $20$ candidate programs and,
instead of refining the specification, select one using two common
approaches: CodeT~\cite{chen2023codet} and majority voting. CodeT first clusters
the candidates by their outputs on the public inputs, then selects a
program from the largest cluster. Majority voting selects the program
that agrees most often with the majority output on each public input.

Both alternatives perform worse than \sys on all three LLMs.
Averaged over the four benchmarks, relative drops in \pat{} and \apr{}
on \texttt{qwen3-coder-30b} are 25.6\% and 18.0\% for CodeT,
and 24.6\% and 17.0\% for majority voting. The gaps are also substantial on
\texttt{glm-4.7-flash}, reaching 40.3\% and 34.1\% for CodeT,
and 26.9\% and 11.1\% for majority voting. In contrast, the drops
on \texttt{gpt-5-mini} are smaller: 6.1\% and 5.7\% for CodeT,
and 4.7\% and 3.9\% for majority voting.
Detailed experimental settings and per-benchmark results are
available in our replication package.

Thus, candidate generation with simple selection heuristics does not
match \sys, particularly on the open-weight LLMs. A possible explanation is
that multiple candidates can share the same misinterpretation of
the specification. Agreement-based selection then favors
a common answer without correcting the misunderstanding.
Behavior-guided refinement can identify and correct
these defects rather than merely select among implementations. The smaller gaps on \texttt{gpt-5-mini}
suggest that heuristic selection is more competitive on the stronger
model, whereas the open-weight LLMs benefit more from the full
behavior-guided refinement and selection process.

\subsection{How Reliable Are the Predicted Behaviors?}
\label{sec:disc-behaviors}

\sys uses predicted behaviors to identify specification defects
and guide repairs. However, these predictions can themselves be
wrong and lead to incorrect repairs. To examine this risk, we
audited the same 40 problems per LLM from the preceding analysis,
with three repair rounds per problem.
We checked the predicted behaviors using reference solutions or
programs that pass all hidden tests. We also compared pre- and post-repair specifications
against the problem statement to check whether repairs corrected
or introduced errors.
Detailed checking procedures and per-run records are available
in our replication package.

Among the checkable behaviors \sys accepted as required, 0/88 were incorrect for \texttt{gpt-5-mini},
23/104 (22.1\%) for \texttt{glm-4.7-flash}, and 8/83 (9.6\%)
for \texttt{qwen3-coder-30b}. Yet wrong predictions do not
automatically lead to wrong repairs. \sys first checks whether
a behavior is required, then determines whether its violation
comes from a specification defect. Across the 40 runs per model,
we found 18 improvements and 2 wrong fixes for
\texttt{gpt-5-mini}, 16 and 4 for \texttt{glm-4.7-flash},
and 21 and 6 for \texttt{qwen3-coder-30b}. Only one of the
12 wrong-fix runs contained a required behavior we found to be false. The other wrong fixes occurred when diagnosis
or rewriting added a false rule or removed a correct one.
Thus, reliability depends not only on behavior prediction but
also on how the resulting feedback is used to repair the specification.

Candidate agreement can also hide errors when all programs share
the same misinterpretation. We observed this in 10 of the 120 runs.
In all 10, a public example exposed the error and provided feedback
for repair despite the candidates' agreement. Public examples
therefore complement predicted behaviors, but neither source
guarantees that every misunderstanding will be detected.

\subsection{Threats to Validity}
\label{sec:threats}

\emph{Construct validity} concerns whether the metrics used in the evaluation
faithfully capture the concepts of interest. Following prior work on
specification alignment~\cite{specfix,specine,tian2023mufix}, we use \pat{} and \apr{} to measure program
correctness, with \pat{} capturing exact success and \apr{} capturing partial
correctness.  To
provide a broader measurement, we additionally report efficiency metrics,
including token and time cost, and conduct a failure analysis to characterize the
remaining errors. 

\emph{Internal validity} concerns whether the observed improvements are caused by the
method itself rather than by implementation choices, model selection, or
experimental configuration. The choice of LLMs may affect the experimental results,
since different model families have different coding and reasoning capabilities.
We mitigate this threat by using three representative LLM types within our
financial constraints: a coding-oriented open-weight model, a reasoning-oriented open-weight model, and a general closed-source model. Experimental configuration
may also bias the comparison. We mitigate this by using the same evaluation setup
for \sys and the baselines where applicable, including the same benchmark splits,
LLM temperature, iteration limits, candidate pool sizes, and scoring protocol.

\emph{External validity }  concerns whether the results generalize beyond the evaluated dataset. To assess the generalizability, we evaluate \sys on four benchmark datasets and six evaluation configurations, covering different problem sources and test settings. Moreover, older benchmarks may also overlap with the training data of recent LLMs. To mitigate data leakage, we also include the LiveCodeBench dataset, a more recent benchmark. The consistent gains of \sys on LiveCodeBench suggest that the improvements are unlikely to rely solely on memorized benchmark solutions. The higher token and time costs of \sys compared with most baselines may limit its use in resource-constrained settings. Our preliminary observations suggest that lowering the maximum output token limit can reduce both costs with an acceptable performance loss. In practice, we suggest setting this limit according to the problem's needs rather than always using the model's maximum supported output length. A systematic study of this trade-off is left for future work.

\emph{Conclusion validity} concerns whether the evidence is strong enough to support the
claims drawn from the experiments, rather than reflecting random variation or an
isolated benchmark effect. We mitigate this threat by evaluating on $877$ unique
problems across four benchmarks and six evaluation configurations, and by
repeating the evaluation across three LLMs. Across these settings, \sys shows consistent and substantial improvements over the strongest baselines, reducing the likelihood that the conclusions are driven by a single model, benchmark, or random outcome.

\section{Related Work}
\label{sec:related}

\subsection{Automated Code Generation}
Code generation has been studied for decades, and recent advances in large
language models trained on code~\cite{chen2021codex,austin2021program} have
accelerated progress in LLM-based program generation. As discussed in the
introduction, most existing methods strengthen planning, implementation, and repair
rather than intent interpretation. Prompting methods such as structured chain-of-thought~\cite{li2023scot}
improve planning before coding. Generate-then-refine methods iterate on the
implementation using execution feedback or self-feedback, including self-debugging
\cite{chen2023selfdebug}, self-repair~\cite{olausson2023selfrepair}, and
self-refine~\cite{madaan2023selfrefine}. Multi-agent frameworks assign
complementary roles to strengthen planning, implementation, and testing:
Self-Collaboration~\cite{dong2024selfcollab} and AgentCoder
\cite{huang2023agentcoder} coordinate analyst/coder/tester agents, and PairCoder
\cite{zhang2024paircoder} mimics pair programming with a navigator and a driver.
Selection-based methods rerank sampled implementations using generated tests or
consensus, as in CodeT~\cite{chen2023codet} and multi-perspective
self-consistency~\cite{huang2023mpsc}. In contrast, \sys targets intent interpretation.
It refines the LLM's interpretation of the original intent rather than the generated code.

\subsection{Specification Alignment}
A growing line of work recognizes that ambiguous intent, not generation ability,
often limits LLM code generation. $\mu$FiX~\cite{tian2023mufix} improves the
model's specification \emph{understanding} through test-case analysis and feedback
prompting. ClarifyGPT~\cite{mu2024clarifygpt} detects ambiguous requirements and
asks clarifying questions. Specine~\cite{specine} lifts the LLM-perceived
specification out of generated code into a requirement DSL and aligns it to the
input using software-engineering rules. SpecFix~\cite{specfix} repairs the
distribution of programs an ambiguous requirement induces and maps the change back
to the requirement via contrastive specification inference. \sys shares the
premise of these works that specification alignment is central, but differs in
mechanism. Rather than rewriting the requirement and then generating code
directly, \sys maintains an \emph{explicit} model of intended behavior across various input scenarios. It then uses \emph{disagreement among
executed candidates} to localize where the specification is ambiguous and to decide whether a failure comes from the specification or the generated code. Our experiments show that this behavior-level specification alignment outperforms both Specine and SpecFix.

Specification refinement is also relevant to automated program repair.
SpecRover~\cite{ruan2025specrover}, for example, infers specifications to guide
repairs in an existing repository. Our setting is different:
\sys targets code generation from natural-language intent, without
an existing implementation to repair. It generates candidate programs
from scratch and uses discrepancies between predicted and observed
behaviors to identify and refine unclear specification clauses.

\section{Conclusion and Future Work}
\label{sec:conclusion}

In this work, we presented \sys, a framework for specification alignment in LLM code generation. Instead of only refining generated code after failures, \sys builds an explicit behavior model of the correct program from the original intent, checks candidate programs by execution, and uses the gap between predicted and observed behavior to refine the
specification and select the best-aligned program. Across three LLMs and six evaluation settings spanning four benchmarks, including the more recent
LiveCodeBench, \sys consistently improves both \pat{} and \apr{} over nine
baselines. The failure analysis further shows that, after alignment, most
remaining failures are adherence failures, primarily algorithmic/logic errors,
runtime crashes, and timeouts.

In the future, we plan to extend specification alignment beyond generated code. One
direction is to align specifications with additional sources of evidence, such as
tests and execution traces. This would allow the framework to reason jointly
about the problem statement, the generated code, and the available tests,
supporting more reliable LLM-based software development when intent is ambiguous
and evidence is incomplete.

\section*{Acknowledgment}
This work has emanated from research jointly funded by Taighde \'{E}ireann -- Research Ireland under Grant number 13/RC/2094\_2 and by Huawei Technologies Co., Ltd. Lionel Briand is also supported by the Natural Sciences and Engineering Research Council of Canada. For the purpose of Open Access, the authors have applied a CC BY public copyright licence to any Author Accepted Manuscript version arising from this submission.

\bibliographystyle{IEEEtran}
\bibliography{references}
\end{document}